\documentclass[rmp,amsfonts,showpacs,showkeys,preprint]{revtex4}
\usepackage[x11names]{xcolor}
\usepackage{graphicx}
\usepackage{dcolumn}
\usepackage{bm}
\usepackage{eepic}
\bibpunct{[}{]}{,}{a}{}{;}
\RequirePackage{times}
\RequirePackage{mathptm}
\usepackage{longtable}

\begin{document}

\title{Bertlmann's chocolate balls and quantum type cryptography}

\author{Karl Svozil}
\email{svozil@tuwien.ac.at}
\homepage{http://tph.tuwien.ac.at/~svozil}
\affiliation{Institute for Theoretical Physics, Vienna University of Technology,  \\
Wiedner Hauptstra\ss e 8-10/136, A-1040 Vienna, Austria}

\begin{abstract}
Some quantum cryptographic protocols can be implemented with specially prepared chocolate balls, others protected by value indefiniteness cannot. Similarities and differences of cryptography with quanta and chocolate are discussed. Motivated by these considerations it is proposed to certify quantum random number generators and quantum cryptographic protocols by value indefiniteness. This feature, which derives itself from Bell- and Kochen-Specker type arguments, is only present in systems with three or more mutually exclusive outcomes.
\end{abstract}


\maketitle

\section{Quantum resources for cryptography}

Quantum cryptography\footnote{
In view of the many superb presentations of quantum cryptography
--- to name but a few, see Refs.~\cite{gisin-qc-rmp,arXiv:0802.4155}  and
\cite[Chapter~6]{mermin-04} (or, alternatively, \cite[Section 6.2]{mermin-07}),
as well as
\cite[Section~12.6]{nielsen-book};
apologies to other authors for this incomplete, subjective collection
---
we refrain from any extensive introduction.
}
uses quantum resources to encode plain symbols forming some message.
Thereby, the security of the code against cryptanalytic attacks to recover
that message rests upon the validity of physics, giving new and direct meaning to
Landauer's dictum~\cite{landauer} ``information is physical.''

What exactly are those quantum resources on which quantum cryptography is based upon?
Consider, for a start,  the following qualities of quantized systems:
\renewcommand{\labelenumi}{(\roman{enumi})}
\begin{enumerate}
\item
randomness of certain individual events,
such as the occurrence of certain measurement outcomes
for states which are in a  superposition of eigenstates
associated with eigenvalues corresponding to these outcomes;
\item
complementarity, as proposed by Pauli, Heisenberg and Bohr;
\item
value indefiniteness, as attested by Bell, Kochen \& Specker and others
(often, this property is referred to as ``contextuality'');
\item
interference and quantum parallelism, allowing the co-representation of classically contradicting states of information
by a coherent superposition thereof;
\item
entanglement of two or more particles,
as pointed out by Schr\"odinger, such that their state cannot be represented
as the product of states of the isolated, individual quanta,
but is rather defined by the {\em joint} or {\em relative} properties of the quanta involved.
\end{enumerate}

The first quantum cryptographic protocols, such as the ones by Wiesner~\cite{wiesner} and
Bennett \& Brassard~\cite{benn-84,benn-92},
just require complementarity and random individual outcomes.
This might be perceived ambivalently as and advantage --- by being based upon only these two features ---
yet also as a disadvantage, since they are not ``protected'' by Bell- or Kochen-Specker type
value indefiniteness.

This article addresses two issues: a critical re-evaluation
of quantum cryptographic protocols in view
of quantum value indefiniteness;
as well as suggestions to improve them to assure the best possible protection
``our''~\cite[p.~866]{born-26-1} present quantum theory can afford.
In doing so, a toy model will be introduced which implements complementarity but still
is value definite.
Then it will be exemplified how to do perform ``quasi-classical'' quantum-like cryptography
with these models.
Finally, methods will be discussed which go beyond the quasi-classical realm.

Even nowadays it is seldom acknowledged that,
when it comes to value definiteness, there definitely {\em is} a difference between
two- and three-dimensional Hilbert space.
This difference can probably be best explained in terms of (conjugate) bases:
whereas different basis in two-dimensional Hilbert space are disjoint and separated
(they merely share the trivial origin),
from three dimensions onwards, they may share common elements.
It is this inter-connectedness of bases and ``frames'' which
supports both Gleason's and the Kochen-Specker theorem.
This can, for instance, be used in derivations of the latter one in three dimensions,
which effectively amount to a succession of rotations of bases along one of their elements
(the original Kochen-Specker~\cite{kochen1} proof uses 117 interlinked bases), thereby creating new rotated bases,
until the original base is reached.
Note that certain (even dense~\cite{meyer:99}) ``dilutions'' of bases break up the possibility to interconnect,
thus allowing value definiteness.

The importance of these arguments for physics is this:
since in quantum mechanics the dimension of Hilbert space
is determined by the number of mutually exclusive outcomes,
a {\em necessary} condition for a quantum system to be protected by value indefiniteness
thus is that the associated quantum system has {\em at least three} mutually exclusive outcomes;
two outcomes are insufficient for this purpose.
Of course, one could argue that systems with two outcomes are still protected by complementarity.

\section{Realizations of quantum cryptographic protocols}

Let us, for the sake of demonstration,
discuss a concrete ``toy'' system which features complementarity but (not) value (in)definiteness.
It is based on the partitions of a set.
Suppose the set is given by
$S=\{1,2,3,4\}$,
and consider two of its equipartitions
$A=\{\{1,2\},\{3,4\}\}$
and
$B=\{\{1,3\},\{2,4\}\}$, as well as
the usual set theoretic operations (intersection, union and complement) and
the subset relation among the elements of these two partitions.
Then $A$ and $B$ generate two Boolean algebras
$L_A= \{\emptyset ,\{1,2\},\{3,4\},S\}$
and
$L_B= \{\emptyset ,\{1,3\},\{2,4\},S\}$
which are equivalent to $2^2$; with two atoms
$a_1=\{1,2\}$ \& $a_2=\{3,4\}$, as well as
$b_1=\{1,3\}$  \& $b_2=\{2,4\}$
per algebra, respectively.
Then, the partition logic
$L_A \oplus L_B = L_{A,B} = \left\langle \{L_A,L_B\},\cap, \cup, ',\subset \right\rangle $
is obtained as a pasting construction from $L_A$ and $L_B$:
only elements contribute which are in $L_A$, or in $L_B$, or in both $L_A \cap L_B$ of them
(the atoms of this algebra being the elements  $a_1,\ldots ,b_2$),
and all common elements --- in this case only the smallest and greatest elements $\emptyset$
and $S$  ---  are identified.
$L_{A,B}$ ``inherits'' the operations and relations of its subalgebras (also called {\em blocks}
or {\em contexts}) $L_A$ and $L_B$.
This pasting construction yields a nondistributive and thus
nonboolean, orthocomplemented propositional structure.
Nondistributivity can quite easily be proven,
as  $a_1 \wedge (b_1 \vee b_2) \neq  (a_1 \wedge b_1) \vee (a_1 \wedge b_2)$,
since $b_1 \vee b_2=S$, whereas  $a_1 \wedge b_1= a_1 \wedge b_2 =\emptyset$.
Note that, although $a_1,\ldots ,b_2$ are compositions of elements of $S$,
not all elements of the power set $2^S\equiv 2^4$ of $S$,  such
as $\{1\}$ or $\{1,2,3\}$, are contained in $L_{A,B}$.

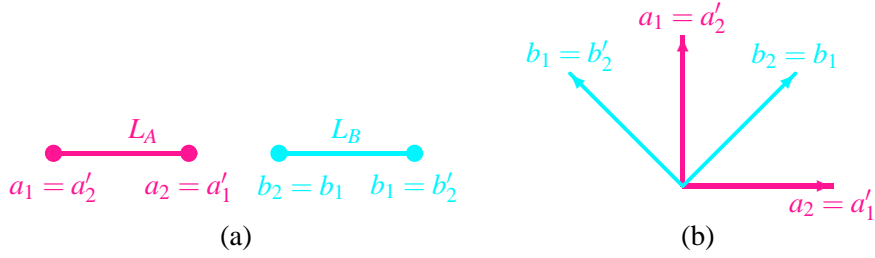
\begin{figure}
\begin{center}
\begin{tabular}{ccc}
\unitlength 0.30mm
\allinethickness{1.5pt} 
\begin{picture}(165,65.00)
\put(0.00,15.00){\color{DeepPink1}\circle{6}}
\put(0.00,15.00){\color{DeepPink1}\circle{2.8}}
\put(0.00,15.00){\color{DeepPink1}\circle{0.5}}
\put(60.00,15.00){\color{DeepPink1}\circle{6}}
\put(60.00,15.00){\color{DeepPink1}\circle{2.8}}
\put(60.00,15.00){\color{DeepPink1}\circle{0.5}}
\put(0.00,0.00){\color{DeepPink1}\makebox(0,0)[cc]{$a_1=a_2'$}}
\put(60.00,0.00){\color{DeepPink1}\makebox(0,0)[cc]{$a_2=a_1'$}}
\put(0.00,15.00){\color{DeepPink1}\line(1,0){60.00}}
\put(100.00,15.00){\color{Turquoise1}\circle{6}}
\put(100.00,15.00){\color{Turquoise1}\circle{2.8}}
\put(100.00,15.00){\color{Turquoise1}\circle{0.5}}
\put(160.00,15.00){\color{Turquoise1}\circle{6}}
\put(160.00,15.00){\color{Turquoise1}\circle{2.8}}
\put(160.00,15.00){\color{Turquoise1}\circle{0.5}}
\put(110.00,0.00){\color{Turquoise1}\makebox(0,0)[cc]{$b_2=b_1$}}
\put(160.00,0.00){\color{Turquoise1}\makebox(0,0)[cc]{$b_1=b_2'$}}
\put(100.00,15.00){\color{Turquoise1}\line(1,0){60.00}}
\put(40.00,25.00){\color{DeepPink1}\makebox(0,0)[cc]{$L_A$}}
\put(130.00,25.00){\color{Turquoise1}\makebox(0,0)[cc]{$L_B$}}
\end{picture}
&
\qquad
\qquad
\qquad
&
\unitlength .5mm 
\allinethickness{1.5pt} 
\ifx\plotpoint\undefined\newsavebox{\plotpoint}\fi 
\begin{picture}(69.971,40.225)(0,0)
\put(29.971,.225){\color{DeepPink1}\vector(0,1){40}}
\put(29.971,.225){\color{DeepPink1}\vector(1,0){40}}
\put(29.971,.225){\color{Turquoise1}\vector(1,1){30}}
\put(29.943,.225){\color{Turquoise1}\vector(-1,1){30}}
\put(30,45){\color{DeepPink1}\makebox(0,0)[cc]{$a_1=a_2'$}}
\put(69.971,-4.775){\color{DeepPink1}\makebox(0,0)[cc]{$a_2=a_1'$}}
\put(60,35){\color{Turquoise1}\makebox(0,0)[cc]{$b_2=b_1$}}
\put(0,35){\color{Turquoise1}\makebox(0,0)[]{$b_1=b_2'$}}
\end{picture}
\\
(a)&&(b)
\end{tabular}
\end{center}
\caption{\label{2006-ql-nondist}
(a) Greechie
diagram  of $L_{A,B}$, consisting of two separate Boolean subalgebras $L_A$ and $L_B$;
(b)
two-dimensional configuration of spin-$\frac{1}{2}$
state measurements along two noncollinear directions.
As there are only two mutually exclusive outcomes, the dimension of the Hilbert space is two.
}
\end{figure}

Figure~\ref{2006-ql-nondist}(a)
depicts a Greechie (orthogonality) diagram of  $L_{A,B}$,
which represents elements in a Boolean algebra as single smooth curves;
in this case there are just two atoms  (least elements above $\emptyset$)
per subalgebra; and both subalgebras are not interconnected.

Several realizations of this partition logic exist; among them
\begin{enumerate}
\item
the propositional structure~\cite{birkhoff-36,svozil-ql} of spin state measurements of a spin-$\frac{1}{2}$ particle
along two noncollinear directions,
or of the linear polarization of a photon along two nonorthogonal, noncollinear directions.
A  two-dimensional Hilbert space representation of this configuration is depicted in
Figure~\ref{2006-ql-nondist}(b).
Thereby, the choice of the measurement direction decides which one of the two
complementary spin state observables is measured;
\item
generalized urn models~\cite{wright,dvur-pul-svo};
in particular ones with black balls painted
with two symbols having two possible values (say, ``$0$´´ and ``$1$´´)
in two colors (say,  ``red'' and ``green''),
resulting in four types of balls
---
more explicitly, carrying all variation of the symbols
\unitlength 0.7mm \allinethickness{1pt}\begin{picture}(8,8) \put(4,2){\circle*{8}} \put(4,2){\makebox(0,0)[cc]{${\color{DeepPink1} {\bf 0}}{\color{Turquoise1} {\bf 0}}$}} \end{picture},
\unitlength 0.7mm \allinethickness{1pt}\begin{picture}(8,8) \put(4,2){\circle*{8}} \put(4,2){\makebox(0,0)[cc]{${\color{DeepPink1} {\bf 0}}{\color{Turquoise1} {\bf 1}}$}} \end{picture},
\unitlength 0.7mm \allinethickness{1pt}\begin{picture}(8,8) \put(4,2){\circle*{8}} \put(4,2){\makebox(0,0)[cc]{${\color{DeepPink1} {\bf 1}}{\color{Turquoise1} {\bf 0}}$}} \end{picture}, as well as
\unitlength 0.7mm \allinethickness{1pt}\begin{picture}(8,8) \put(4,2){\circle*{8}} \put(4,2){\makebox(0,0)[cc]{${\color{DeepPink1} {\bf 1}}{\color{Turquoise1} {\bf 1}}$}} \end{picture}
--- many copies of which are randomly
distributed in an urn.
Suppose the experimenter looks at them with one of two differently colored eyeglasses,
each one ideally  matching the colors of only one of the symbols,
such that only light in this wave length passes through.
Thereby, the choice of the color decides which one of the two
complementary observables associated with ``red'' and ``green'' is measured.
Propositions refers to the possible ball types drawn from the urn, given the information printed in the chosen color.
\item
initial state identification problem for
deterministic finite (Moore or Mealy)
automata in an unknown initial state~\cite{e-f-moore,svozil-2001-eua};
in particular ones $\left\langle S, I, O, \delta, \lambda \right\rangle$
with four internal states $S=\{1,2,3,4\}$, two input and two output states $I=O=\{0,1\}$,
 an ``irreversible'' (all-to-one) transition function
$\delta (s, i) =1$ for all $s\in S$, $i\in I$,
and an output function  ``modelling'' the state partitions by
$\lambda (1,0)=\lambda (2,0) =0$,
$\lambda (3,0)=\lambda (4,0) =1$,
$\lambda (1,1)=\lambda (3,1) =0$,
$\lambda (2,1)=\lambda (4,1) =1$.
Thereby, the choice of the input symbol decides which one of the two
complementary observables is measured.
\end{enumerate}

Let us, for the moment, consider generalized urn models,
because they allow a ``pleasant'' representation as chocolate balls coated in black foils and
painted with color symbols.
With the four types of chocolate balls
\unitlength 0.7mm \allinethickness{1pt}\begin{picture}(8,8) \put(4,2){\circle*{8}} \put(4,2){\makebox(0,0)[cc]{${\color{DeepPink1} {\bf 0}}{\color{Turquoise1} {\bf 0}}$}} \end{picture},
\unitlength 0.7mm \allinethickness{1pt}\begin{picture}(8,8) \put(4,2){\circle*{8}} \put(4,2){\makebox(0,0)[cc]{${\color{DeepPink1} {\bf 0}}{\color{Turquoise1} {\bf 1}}$}} \end{picture},
\unitlength 0.7mm \allinethickness{1pt}\begin{picture}(8,8) \put(4,2){\circle*{8}} \put(4,2){\makebox(0,0)[cc]{${\color{DeepPink1} {\bf 1}}{\color{Turquoise1} {\bf 0}}$}} \end{picture}, and
\unitlength 0.7mm \allinethickness{1pt}\begin{picture}(8,8) \put(4,2){\circle*{8}} \put(4,2){\makebox(0,0)[cc]{${\color{DeepPink1} {\bf 1}}{\color{Turquoise1} {\bf 1}}$}} \end{picture}
drawn from an urn it is
possible to execute the 1984 Bennett-Brassard (BB84) protocol~\cite{benn-84,benn-92}
and ``generate'' a secret key shared by two parties~\cite{svozil-2005-ln1e}.
Formally, this reflects
(i) the random draw of balls from an urn, as well as
(ii) the complementarity modeled
{\em via} the color painting and the colored eyeglasses.
It also reflects the possibility to embed this model into a bigger Boolean (and thus classical)
algebra $2^4$ by ``taking off the eyeglasses'' and looking at both symbols of those four balls types
simultaneously.
The atoms of this Boolean algebra are just the ball types, associated with the four  cases
\unitlength 0.7mm \allinethickness{1pt}\begin{picture}(8,8) \put(4,2){\circle*{8}} \put(4,2){\makebox(0,0)[cc]{${\color{DeepPink1} {\bf 0}}{\color{Turquoise1} {\bf 0}}$}} \end{picture},
\unitlength 0.7mm \allinethickness{1pt}\begin{picture}(8,8) \put(4,2){\circle*{8}} \put(4,2){\makebox(0,0)[cc]{${\color{DeepPink1} {\bf 0}}{\color{Turquoise1} {\bf 1}}$}} \end{picture},
\unitlength 0.7mm \allinethickness{1pt}\begin{picture}(8,8) \put(4,2){\circle*{8}} \put(4,2){\makebox(0,0)[cc]{${\color{DeepPink1} {\bf 1}}{\color{Turquoise1} {\bf 0}}$}} \end{picture}, and
\unitlength 0.7mm \allinethickness{1pt}\begin{picture}(8,8) \put(4,2){\circle*{8}} \put(4,2){\makebox(0,0)[cc]{${\color{DeepPink1} {\bf 1}}{\color{Turquoise1} {\bf 1}}$}} \end{picture}.
The possibility of a classical embedding is also reflected in a ``sufficient'' number
(i.e., by a separating, full set) of two-valued, dispersionless states
$P(a_1)+P(a_2)=P(b_1)+P(b_2)=1$, with $P(x)\in \{0,1\}$.
These two-valued states can also be interpreted as logical truth assignments,
irrespective of whether or not the observables have been (co-)measured.

The possibility to ascribe certain ``ontic states'' interpretable as observer-independent
``omniscient elements of physical reality''
(in the sense of Einstein, Podolsky and Rosen~\cite[p.~777]{epr},
a paper which amazingly contains not a single reference) even for complementarity observables
may raise some skepticism or even outright rejection,
since that is not how quantum mechanics is known to perform ``at its most mind-boggling mode.''
Indeed, so far, the rant presented merely attempted to convince the reader that one can
have complementarity {\em as well as} value definiteness; i.e.,
complementarity is not sufficient for value indefiniteness in the
sense of the Bell- and Kochen-Specker argument.

Unfortunately, the two-dimensionality of the associated Hilbert space
is also a feature
plaguing present random number generators based on beam
splitters~\cite{svozil-qct,rarity-94,zeilinger:qct,stefanov-2000}.
In this respect, most of the present random number generators using
beam splitters are protected only by the randomness
of single outcomes as well as by complementarity, but are not by
certified value indefiniteness,
as guaranteed by quantum theory in its standard form~\cite{v-neumann-49}.
Their methodology should also be improved by the methods discussed below.

\section{Supporting cryptography with value indefiniteness}

Alas, quantum mechanics is more resourceful and mind-boggling than that,
as it does not permit any two-valued states which may be ontologically  interpretable
as elements of physical reality.
So we have to go further, reminding ourselves that value indefiniteness
comes about only for Hilbert spaces of dimensions three and higher.
There are several ways of doing this.
The following options will be discussed:
\begin{enumerate}
\item
the known protocols can be generalized to three or more outcomes;
\item
entangled pairs of particles~\cite{ekert91} associated with statistical value indefiniteness may be considered;
\item
full, nonprobabilistic value indefiniteness may be attempted, alt least counterfactually.
\end{enumerate}

\subsection{Generalizations to three and more outcomes}

In constructing quantum random number generators {\em via} beam splitters
which ultimately are used in cryptographic setups, it is important
(i)  to have full control of the particle source, and
(ii) to use beam splitters with three or more output ports,
associated with three- or higher-dimensional Hilbert spaces.
Thereby, it is {\em not sufficient} to compose a multiport beam splitter
by a succession of phase shifters and beam splitters
with two output ports~\cite{rzbb,svozil-2004-analog},
based on elementary decompositions of the unitary group~\cite{murnaghan}.

Dichotomic sequences could be obtained from sequences containing more than two symbols
by discarding all other symbols from that sequence~\cite{MR997340},
or by identifying the additional symbols with one (or both) of the two symbols.
For standard normalization procedures and their issues,
the reader is referred to Refs.~\cite{von-neumann1,elias-72,PeresY-1992,dichtl-2007,Lacharme-2008}.

One concrete realization would be a spin-$\frac{3}{2}$ particle. Suppose it is prepared
in one of its four spin states, say the one associated with angular momentum $+\frac{3}{2}\hbar$ in
some arbitrary but definite direction; e.g., by a Stern-Gerlach device.
Then, its  spin state is  again measured along a perpendicular direction;
e.g., by another, differently oriented, Stern-Gerlach device.
Two of the output ports, say the ones corresponding to positive angular momentum $+\frac{3}{2}\hbar$
and $+\frac{1}{2}\hbar$,
are identified with the symbol ``$0$,'' the other two ports with the symbol ``$1$.''
In that way, a random sequence is obtained from quantum coin tosses
which can be ensured to operate under the conditions of value indefiniteness
in the sense of the Kochen-Specker theorem.
Of course, this protocol can also be used to generate random sequences
containing four symbols (one symbol per detector).

With respect to the use of beam splitters, the reader is kindly reminded of another
issue related to the fact that beam splitters are {\em reversible} devices capable of only
translating an incoming  signal into an outgoing  signal in a {\em one-to-one} manner.
The ``nondestructive'' action of a beam splitter could also be demonstrated
by ``reconstructing'' the original signal through a ``reversed'' identical beam splitter
in a Mach-Zehnder interferometer~\cite{green-horn-zei}.
In this sense, the signal leaving the output ports of a beam splitter
is ``as good'' for cryptographic purposes as the one entering the device.
This fact relegates considerations of the quality of quantum randomness
to the quality of the source. Every care should thus be taken in preparing the source
to  assure that the state entering the input port
(i)  either is pure and could subsequently be used for measurements corresponding to conjugate bases,
(ii) or is maximally mixed, resulting in a representation of its
state in finite dimensions proportional to the unit matrix.

\subsection{Configurations with statistical value indefiniteness}

Protocols like the Ekert protocol \cite{ekert91}
utilize two entangled two-state particles for a generation of
a random key shared by two parties. The particular Einstein-Podolsky-Rosen
configuration~\cite{epr} and the singlet Bell state communicated among the
parties guarantee  stronger-than-classical correlations of their sequences,
resulting in a violation of Bell-type inequalities obeyed by classical probabilities.

Although criticized~\cite{PhysRevLett.68.557} on the grounds that the Ekert protocol
in certain cryptanalytic aspects is equivalent to existing ones
(see Ref.~\cite{benn-92b} for a reconciliation),
it offers additional security in the light of quantum value indefiniteness,
as it suggests to probe the nonclassical parts of quantum statistics.
This can best be understood in terms of the impossibility to generate
co-existing tables of all --- even the counterfactually possible --- measurement outcomes
of the quantum observables used~\cite{peres222}.
This, of course, can only happen for the four-dimensional Hilbert space configuration
proposed by Ekert,
and not for effectively two-dimensional ones of previous proposals.
As a result, the Eckert protocol cannot be performed with chocolate balls.
Formally, this is due to the nonexistence of two-valued states
in four-dimensional Hilbert space.

Suppose one would nevertheless attempt to ``mimic'' the Ekert protocol with a classical ``singlet'' state
which uses compositions of two balls of the form
\unitlength 0.7mm \allinethickness{1pt}\begin{picture}(8,8) \put(4,2){\circle*{8}} \put(4,2){\makebox(0,0)[cc]{${\color{DeepPink1} {\bf 0}}{\color{Turquoise1} {\bf 0}}$}} \end{picture}---\unitlength 0.7mm \allinethickness{1pt}\begin{picture}(8,8) \put(4,2){\circle*{8}} \put(4,2){\makebox(0,0)[cc]{${\color{DeepPink1} {\bf 1}}{\color{Turquoise1} {\bf 1}}$}} \end{picture} / \unitlength 0.7mm \allinethickness{1pt}\begin{picture}(8,8) \put(4,2){\circle*{8}} \put(4,2){\makebox(0,0)[cc]{${\color{DeepPink1} {\bf 0}}{\color{Turquoise1} {\bf 1}}$}} \end{picture}---\unitlength 0.7mm \allinethickness{1pt}\begin{picture}(8,8) \put(4,2){\circle*{8}} \put(4,2){\makebox(0,0)[cc]{${\color{DeepPink1} {\bf 1}}{\color{Turquoise1} {\bf 0}}$}} \end{picture} / \unitlength 0.7mm \allinethickness{1pt}\begin{picture}(8,8) \put(4,2){\circle*{8}} \put(4,2){\makebox(0,0)[cc]{${\color{DeepPink1} {\bf 1}}{\color{Turquoise1} {\bf 0}}$}} \end{picture}---\unitlength 0.7mm \allinethickness{1pt}\begin{picture}(8,8) \put(4,2){\circle*{8}} \put(4
,2){\makebox(0,0)[cc]{${\color{DeepPink1} {\bf 0}}{\color{Turquoise1} {\bf 1}}$}} \end{picture} / \unitlength 0.7mm \allinethickness{1pt}\begin{picture}(8,8) \put(4,2){\circle*{8}} \put(4,2){\makebox(0,0)[cc]{${\color{DeepPink1} {\bf 1}}{\color{Turquoise1} {\bf 1}}$}} \end{picture}---\unitlength 0.7mm \allinethickness{1pt}\begin{picture}(8,8) \put(4,2){\circle*{8}} \put(4,2){\makebox(0,0)[cc]{${\color{DeepPink1} {\bf 0}}{\color{Turquoise1} {\bf 0}}$}} \end{picture},
with strictly different (alternatively strictly identical) particle types.
The resulting probabilities and expectations would obey the classical Clauser-Horne-Shimony-Holt bounds~\cite{chsh}.
This is due to the fact that generalized urn models have quasi-classical probability distributions
which can be represented as convex combinations of the full set of separable two-valued states on their observables.

\subsection{Nonprobabilistic value indefiniteness}

In an attempt to fully utilize quantum value indefiniteness,
we propose a generalization of the BB84 protocol on a propositional structure which does not
allow any two-valued state.
In principle, this could be any kind of finite configuration of observables in three- and higher-dimensional Hilbert space;
in particular ones which have been proposed
for a proof of the Kochen-Specker theorem.
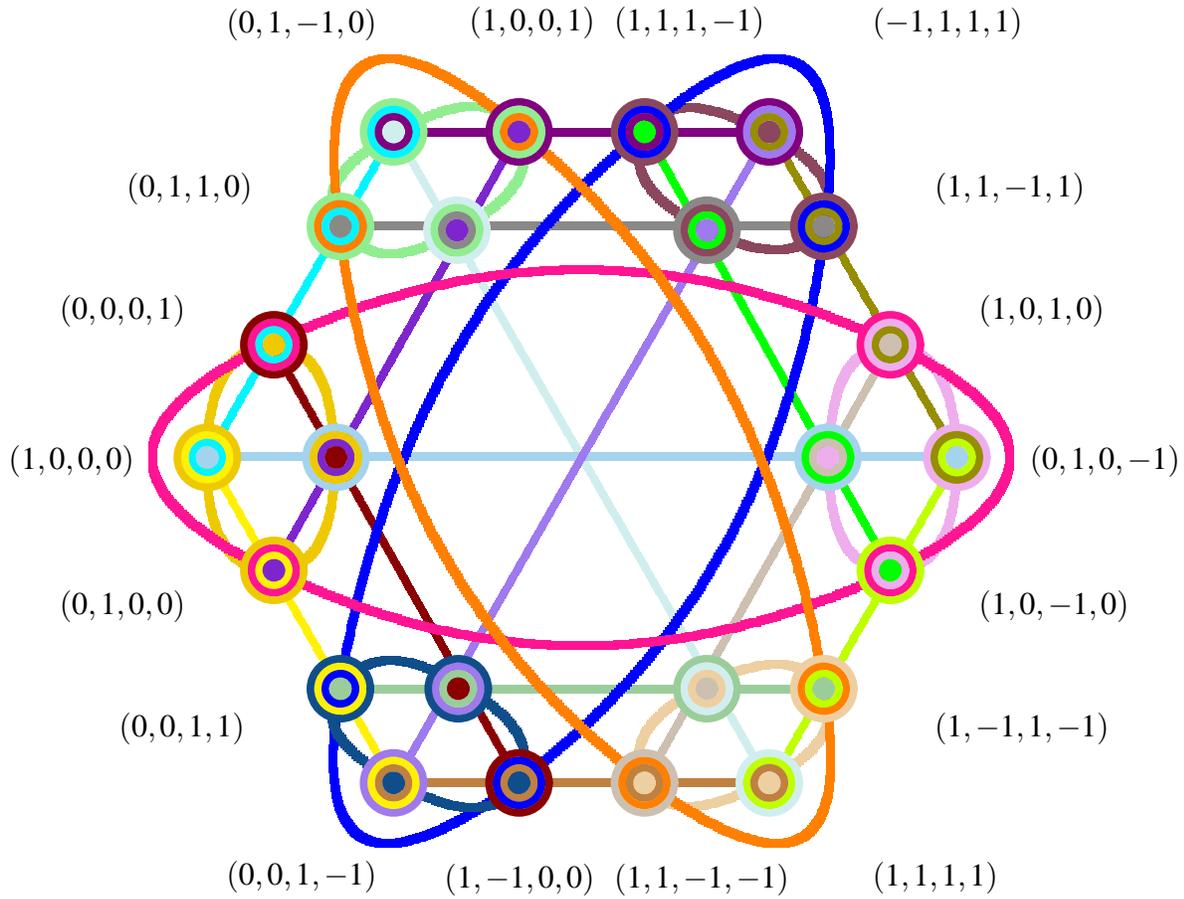
\begin{figure}
\begin{center}

\unitlength 1mm 
\allinethickness{3pt}
\ifx\plotpoint\undefined\newsavebox{\plotpoint}\fi 
\begin{picture}(134.09,125.99)(0,0)

\put(29.168,27.577){{\color{DarkSeaGreen3}\line(1,0){64.169}}}
\multiput(69.826,15.379)(.04816494845,.08565537555){679}{{\color{AntiqueWhite3}\line(0,1){.08565537555}}}
\multiput(102.353,43.487)(-.04816494845,.08617378498){679}{{\color{green}\line(0,1){.08617378498}} }
\put(93.691,89.272){{\color{Snow4}\line(-1,0){64.7}}}
\multiput(53.033,101.823)(-.04816494845,-.08591458027){679}{{\color{Purple3}\line(0,-1){.08591458027}}}
\multiput(20.152,73.539)(.0481420205,-.0854114202){683}{{\color{Red4}\line(0,-1){.0854114202}}}
\multiput(36.592,101.646)(.04819668939,-.08399902629){1027}{{\color{LightCyan2}\line(0,-1){.08399902629}}}
\put(111.015,58.513){{\color{LightSkyBlue2}\line(-1,0){99.348}}}
\multiput(36.416,15.026)(.04816425121,.08386183575){1035}{{\color{MediumPurple2}\line(0,1){.08386183575}}}

\multiput(86.39,14.96)(.119617225,.208133971){209}{{\color{lime}\line(0,1){.208133971}}}
\multiput(86.39,101.96)(.119617225,-.208133971){209}{{\color{olive}\line(0,-1){.208133971}}}
\put(86.39,101.71){{\color{violet}\line(-1,0){50}}}
\multiput(36.47,101.96)(-.119617225,-.208133971){209}{{\color{Turquoise1}\line(0,-1){.208133971}}}
\multiput(36.47,14.96)(-.119617225,.208133971){209}{{\color{yellow}\line(0,1){.208133971}}}
\put(86.39,15.21){{\color{brown}\line(-1,0){50}}}

{\color{DodgerBlue4}\qbezier(52.856,15.203)(56.568,22.009)(44.724,28.461)}{\color{DodgerBlue4}\qbezier(29.345,27.93)(34.029,34.736)(44.724,28.461)}{\color{PaleGreen2}\qbezier(28.991,89.448)(25.279,95.37)(36.416,101.646)}{\color{PaleGreen2}\qbezier(29.345,89.095)(34.029,82.289)(44.724,88.564)}{\color{PaleGreen2}\qbezier(52.856,101.823)(56.568,95.017)(44.724,88.564)}{\color{PaleGreen2}\qbezier(36.416,101.646)(49.055,108.628)(52.856,101.823)}{\color{PaleVioletRed4}\qbezier(69.826,101.646)(73.273,108.452)(86.266,101.469)}{\color{PaleVioletRed4}\qbezier(86.266,101.469)(96.873,95.812)(93.337,89.448)}{\color{PaleVioletRed4}\qbezier(93.337,89.448)(89.713,83.084)(77.958,88.741)}{\color{PaleVioletRed4}\qbezier(77.958,88.741)(66.114,94.663)(69.826,101.646)}{\color{NavajoWhite2}\qbezier(86.266,15.556)(96.873,21.213)(93.337,27.577)}{\color{NavajoWhite2}\qbezier(77.958,28.284)(66.114,22.362)(69.826,15.379)}{\color{NavajoWhite2}\qbezier(93.337,27.577)(89.713,33.941)(77.958,28.284)}{\color{NavajoWhite2}\qbezier(69.826,15.379)
(73.273,8.574)(86.266,15.556)}{\color{Gold2}\qbezier(20.152,73.185)(11.756,73.097)(11.49,58.513)}{\color{Gold2}\qbezier(28.461,58.159)(28.372,73.273)(20.152,73.185)}{\color{Gold2}\qbezier(11.49,58.513)(11.314,43.398)(20.329,43.487)}{\color{Gold2}\qbezier(20.329,43.487)(28.638,43.575)(28.461,58.159)}{\color{Plum2}\qbezier(111.192,58.513)(111.368,43.398)(102.353,43.487)}{\color{Plum2}\qbezier(102.353,43.487)(94.044,43.575)(94.221,58.159)}{\color{Plum2}\qbezier(94.221,58.159)(94.31,73.273)(102.53,73.185)}{\color{Plum2}\qbezier(102.53,73.185)(110.926,73.097)(111.192,58.513)}{\color{blue}\qbezier(29.2,27.73)(23.55,-5.86)(52.99,15.24)}{\color{blue}\qbezier(29.2,27.88)(36.93,75)(69.63,101.91)}{\color{blue}\qbezier(52.69,15.24)(87.47,40.96)(93.72,89.27)}{\color{blue}\qbezier(93.72,89.27)(98.4,125.99)(69.49,102.06)}{\color{DeepPink1}\qbezier(20.15,73.72)(-11.67,58.52)(20.15,43.31)}{\color{DeepPink1}\qbezier(20.33,73.72)(61.34,93.16)(102.36,73.72)}{\color{DeepPink1}\qbezier(102.36,73.72)(134.09,58.52)
(102.53,43.31)}{\color{DeepPink1}\qbezier(102.53,43.31)(60.99,23.43)(20.15,43.49)}{\color{orange}\qbezier(93.57,27.73)(99.22,-5.86)(69.78,15.24)}{\color{orange}\qbezier(93.57,27.88)(85.84,75)(53.13,101.91)}{\color{orange}\qbezier(70.08,15.24)(35.3,40.96)(29.05,89.27)}{\color{orange}\qbezier(29.05,89.27)(24.37,125.99)(53.28,102.06)}{\color{DodgerBlue4}\qbezier(28.991,27.577)(25.279,21.655)(36.416,15.379)}{\color{DodgerBlue4}\qbezier(36.416,15.379)(49.055,8.397)(52.856,15.203)}

\put(11.56,58.45){{\color{Turquoise1}\circle{4}}}
\put(11.56,58.45){{\color{yellow}\circle{6}}}
\put(11.56,58.45){{\color{Gold2}\circle{8}}}
\put(11.56,58.45){{\color{LightSkyBlue2}\circle{2}}}
\put(11.56,58.45){{\color{LightSkyBlue2}\circle{0.1}}}

\put(20.4,73.47){{\color{Turquoise1}\circle{4}}}
\put(20.4,73.47){{\color{DeepPink1}\circle{6}}}
\put(20.4,73.47){{\color{Red4}\circle{8}}}
\put(20.4,73.47){{\color{Gold2}\circle{2}}}
\put(20.4,73.47){{\color{Gold2}\circle{0.1}}}

\put(20.4,43.44){{\color{yellow}\circle{4}}}
\put(20.4,43.44){{\color{DeepPink1}\circle{6}}}
\put(20.4,43.44){{\color{Gold2}\circle{8}}}
\put(20.4,43.44){{\color{Purple3}\circle{2}}}
\put(20.4,43.44){{\color{Purple3}\circle{0.1}}}

\put(28.656,58.45){{\color{Purple3}\circle{4}}}
\put(28.656,58.45){{\color{LightSkyBlue2}\circle{8}}}
\put(28.656,58.45){{\color{Gold2}\circle{6}}}
\put(28.656,58.45){{\color{Red4}\circle{2}}}
\put(28.656,58.45){{\color{Red4}\circle{0.1}}}

\put(94.063,58.45){{\color{LightSkyBlue2}\circle{8}}}
\put(94.063,58.45){{\color{green}\circle{6}}}
\put(94.063,58.45){{\color{AntiqueWhite3}\circle{4}}}
\put(94.063,58.45){{\color{Plum2}\circle{2}}}
\put(94.063,58.45){{\color{Plum2}\circle{0.1}}}

\put(111.21,58.45){{\color{Plum2}\circle{8}}}
\put(111.21,58.45){{\color{olive}\circle{6}}}
\put(111.21,58.45){{\color{lime}\circle{4}}}
\put(111.21,58.45){{\color{LightSkyBlue2}\circle{2}}}
\put(111.21,58.45){{\color{LightSkyBlue2}\circle{0.1}}}

\put(102.37,73.47){{\color{olive}\circle{4}}}
\put(102.37,73.47){{\color{Plum2}\circle{6}}}
\put(102.37,73.47){{\color{DeepPink1}\circle{8}}}
\put(102.37,73.47){{\color{AntiqueWhite3}\circle{2}}}
\put(102.37,73.47){{\color{AntiqueWhite3}\circle{0.1}}}

\put(102.37,43.44){{\color{DeepPink1}\circle{6}}}
\put(102.37,43.44){{\color{lime}\circle{8}}}
\put(102.37,43.44){{\color{Plum2}\circle{4}}}
\put(102.37,43.44){{\color{green}\circle{2}}}
\put(102.37,43.44){{\color{green}\circle{0.1}}}

\put(36.34,15.16){{\color{brown}\circle{4}}}
\put(36.34,15.16){{\color{yellow}\circle{6}}}
\put(36.34,15.16){{\color{MediumPurple2}\circle{8}}}
\put(36.34,15.16){{\color{DodgerBlue4}\circle{2}}}
\put(36.34,15.16){{\color{DodgerBlue4}\circle{0.1}}}

\put(44.919,27.71){{\color{MediumPurple2}\circle{6}}}
\put(44.919,27.71){{\color{DodgerBlue4}\circle{8}}}
\put(44.919,27.71){{\color{DarkSeaGreen3}\circle{4}}}
\put(44.919,27.71){{\color{Red4}\circle{2}}}
\put(44.919,27.71){{\color{Red4}\circle{0.1}}}

\put(29.24,27.71){{\color{yellow}\circle{6}}}
\put(29.24,27.71){{\color{blue}\circle{4}}}
\put(29.24,27.71){{\color{DodgerBlue4}\circle{8}}}
\put(29.24,27.71){{\color{DarkSeaGreen3}\circle{2}}}
\put(29.24,27.71){{\color{DarkSeaGreen3}\circle{0.1}}}

\put(77.976,27.71){{\color{DarkSeaGreen3}\circle{8}}}
\put(77.976,27.71){{\color{LightCyan2}\circle{6}}}
\put(77.976,27.71){{\color{NavajoWhite2}\circle{4}}}
\put(77.976,27.71){{\color{AntiqueWhite3}\circle{2}}}
\put(77.976,27.71){{\color{AntiqueWhite3}\circle{0.1}}}

\put(93.53,27.71){{\color{orange}\circle{6}}}
\put(93.53,27.71){{\color{lime}\circle{4}}}
\put(93.53,27.71){{\color{NavajoWhite2}\circle{8}}}
\put(93.53,27.71){{\color{DarkSeaGreen3}\circle{2}}}
\put(93.53,27.71){{\color{DarkSeaGreen3}\circle{0.1}}}

\put(86.28,15.16){{\color{lime}\circle{6}}}
\put(86.28,15.16){{\color{brown}\circle{4}}}
\put(86.28,15.16){{\color{LightCyan2}\circle{8}}}
\put(86.28,15.16){{\color{NavajoWhite2}\circle{2}}}
\put(86.28,15.16){{\color{NavajoWhite2}\circle{0.1}}}

\put(52.99,15.16){{\color{brown}\circle{4}}}
\put(52.99,15.16){{\color{blue}\circle{6}}}
\put(52.99,15.16){{\color{Red4}\circle{8}}}
\put(52.99,15.16){{\color{DodgerBlue4}\circle{2}}}
\put(52.99,15.16){{\color{DodgerBlue4}\circle{0.1}}}

\put(69.68,15.16){{\color{brown}\circle{4}}}
\put(69.68,15.16){{\color{orange}\circle{6}}}
\put(69.68,15.16){{\color{AntiqueWhite3}\circle{8}}}
\put(69.68,15.16){{\color{NavajoWhite2}\circle{2}}}
\put(69.68,15.16){{\color{NavajoWhite2}\circle{0.1}}}

\put(93.53,89.21){{\color{olive}\circle{4}}}
\put(93.53,89.21){{\color{blue}\circle{6}}}
\put(93.53,89.21){{\color{PaleVioletRed4}\circle{8}}}
\put(93.53,89.21){{\color{Snow4}\circle{2}}}
\put(93.53,89.21){{\color{Snow4}\circle{0.1}}}

\put(77.976,88.679){{\color{Snow4}\circle{8}}}
\put(77.976,88.679){{\color{PaleVioletRed4}\circle{6}}}
\put(77.976,88.679){{\color{green}\circle{4}}}
\put(77.976,88.679){{\color{MediumPurple2}\circle{2}}}
\put(77.976,88.679){{\color{MediumPurple2}\circle{0.1}}}

\put(86.28,101.76){{\color{violet}\circle{8}}}
\put(86.28,101.76){{\color{MediumPurple2}\circle{6}}}
\put(86.28,101.76){{\color{olive}\circle{4}}}
\put(86.28,101.76){{\color{PaleVioletRed4}\circle{2}}}
\put(86.28,101.76){{\color{PaleVioletRed4}\circle{0.1}}}

\put(69.68,101.76){{\color{PaleVioletRed4}\circle{8}}}
\put(69.68,101.76){{\color{blue}\circle{6}}}
\put(69.68,101.76){{\color{violet}\circle{4}}}
\put(69.68,101.76){{\color{green}\circle{2}}}
\put(69.68,101.76){{\color{green}\circle{0.1}}}

\put(52.99,101.76){{\color{violet}\circle{8}}}
\put(52.99,101.76){{\color{PaleGreen2}\circle{6}}}
\put(52.99,101.76){{\color{orange}\circle{4}}}
\put(52.99,101.76){{\color{Purple3}\circle{2}}}
\put(52.99,101.76){{\color{Purple3}\circle{0.1}}}

\put(36.34,101.76){{\color{PaleGreen2}\circle{8}}}
\put(36.34,101.76){{\color{Turquoise1}\circle{6}}}
\put(36.34,101.76){{\color{violet}\circle{4}}}
\put(36.34,101.76){{\color{LightCyan2}\circle{2}}}
\put(36.34,101.76){{\color{LightCyan2}\circle{0.1}}}

\put(44.743,88.679){{\color{LightCyan2}\circle{8}}}
\put(44.743,88.679){{\color{PaleGreen2}\circle{6}}}
\put(44.743,88.679){{\color{Snow4}\circle{4}}}
\put(44.743,88.679){{\color{Purple3}\circle{2}}}
\put(44.743,88.679){{\color{Purple3}\circle{0.1}}}

\put(29.24,89.21){{\color{PaleGreen2}\circle{8}}}
\put(29.24,89.21){{\color{orange}\circle{6}}}
\put(29.24,89.21){{\color{Turquoise1}\circle{4}}}
\put(29.24,89.21){{\color{Snow4}\circle{2}}}
\put(29.24,89.21){{\color{Snow4}\circle{0.1}}}

{\color{black} \normalsize
\put(34,116.02){\makebox(0,0)[rc]{$(0,1,-1,0)$}}
\put(34,2.65){\makebox(0,0)[rc]{$(0,0,1,-1)$}}
\put(63,116.38){\makebox(0,0)[rc]{$(1,0,0,1)$}}
\put(63,2.3){\makebox(0,0)[rc]{$(1,-1,0,0)$}}
\put(100,116.2){\makebox(0,0)[lc]{$(-1,1,1,1)$}}
\put(100,2.48){\makebox(0,0)[lc]{$(1,1,1,1)$}}
\put(65.65,116.38){\makebox(0,0)[lc]{$(1,1,1,-1)$}}
\put(65.65,2.3){\makebox(0,0)[lc]{$(1,1,-1,-1)$}}
\put(108.24,94.22){\makebox(0,0)[lc]{$(1,1,-1,1)$}}
\put(17.45,94.22){\makebox(0,0)[rc]{$(0,1,1,0)$}}
\put(108.24,22.45){\makebox(0,0)[lc]{$(1,-1,1,-1)$}}
\put(16.45,22.45){\makebox(0,0)[rc]{$(0,0,1,1)$}}
\put(114.13,77.96){\makebox(0,0)[lc]{$(1,0,1,0)$}}
\put(8.55,77.96){\makebox(0,0)[rc]{$(0,0,0,1)$}}
\put(114.13,38.72){\makebox(0,0)[lc]{$(1,0,-1,0)$}}
\put(8.55,38.72){\makebox(0,0)[rc]{$(0,1,0,0)$}}
\put(120.92,57.98){\makebox(0,0)[lc]{$(0,1,0,-1)$}}
\put(1.77,57.98){\makebox(0,0)[rc]{$(1,0,0,0)$}}
}
\end{picture}
\end{center}
\caption{ \label{2009-QvPR}
Greechie orthogonality diagram of a ``short'' proof~\cite{cabello-96,cabello:210401} of the Kochen-Specker theorem
in four dimensions containing 24 propositions in 24 tightly interlinked contexts~\cite{tkadlec-priv}.
The graph cannot be colored by the two colors red (associated with truth)
and green (associated with falsity) such that every context contains exactly one red and three green point.
For the sake of a proof, consider just the six outer lines and the three outer ellipses.
Then in a table containing the points of the contexts as columns
and the enumeration of contexts as rows,
every red point occurs in exactly {\em two} contexts, and
there should be an {\em even} number of red points.
On the other hand, there are nine contexts involved; thus by the rules it follows that there
should be an {\em odd} number (nine) of red points in this table (exactly one per context).
}
\end{figure}

\begin{figure}
\begin{center}
\begin{tabular}{cccccc}
\unitlength 0.5mm 
\allinethickness{3pt}
\ifx\plotpoint\undefined\newsavebox{\plotpoint}\fi 
\begin{picture}(132.5,122)(0,0)
\put(20,20){\color{Turquoise1}\line(1,0){110}}
\multiput(20,20)(.03372164316,.05518087063){1631}{\color{yellow}\line(0,1){.05518087063}}
\multiput(75,110)(.03372164316,-.05518087063){1631}{\color{DeepPink1}\line(0,-1){.05518087063}}
\put(20,20){\color{yellow}\circle{5.5}}
\put(20,20){\color{yellow}\circle{1.5}}
\put(20,20){\color{Turquoise1}\circle{9}}
\put(56.25,20){\color{Turquoise1}\circle{5.5}}
\put(56.25,20){\color{Turquoise1}\circle{1.5}}
\put(92.5,20){\color{Turquoise1}\circle{5.5}}
\put(92.5,20){\color{Turquoise1}\circle{1.5}}
\put(129.75,20){\color{Turquoise1}\circle{5.5}}
\put(129.75,20){\color{Turquoise1}\circle{1.5}}
\put(129.75,20){\color{DeepPink1}\circle{9}}
\put(56.25,79.75){\color{yellow}\circle{5.5}}
\put(56.25,79.75){\color{yellow}\circle{1.5}}
\put(38.75,51.25){\color{yellow}\circle{5.5}}
\put(38.75,51.25){\color{yellow}\circle{1.5}}
\put(74.75,109.75){\color{yellow}\circle{9}}
\put(74.75,109.75){\color{DeepPink1}\circle{5.5}}
\put(74.75,109.75){\color{DeepPink1}\circle{1.5}}
\put(93.75,79.75){\color{DeepPink1}\circle{5.5}}
\put(93.75,79.75){\color{DeepPink1}\circle{1.5}}
\put(111.25,51.25){\color{DeepPink1}\circle{5.5}}
\put(111.25,51.25){\color{DeepPink1}\circle{1.5}}
\put(15,5){\makebox(0,0)[cc]{$(1,0,0,0)$}}
\put(52,5){\makebox(0,0)[cc]{$(0,1,1,0)$}}
\put(92.5,5){\makebox(0,0)[cc]{$(0,1,-1,0)$}}
\put(138,5){\makebox(0,0)[cc]{$(0,0,0,1)$}}
\put(74.75,122){\makebox(0,0)[cc]{$(0,1,0,0)$}}
\put(108,80.25){\makebox(0,0)[lc]{$(1,0,-1,0)$}}
\put(125,52.25){\makebox(0,0)[lc]{$(1,0,1,0)$}}
\put(42.5,80.25){\makebox(0,0)[rc]{$(0,0,-1,1)$}}
\put(25.5,52.25){\makebox(0,0)[rc]{$(0,0,1,1)$}}
\end{picture}
&
\qquad
\qquad
\qquad
\qquad
&
\unitlength 0.5mm 
\allinethickness{3pt}
\ifx\plotpoint\undefined\newsavebox{\plotpoint}\fi 
\begin{picture}(132.5,122)(0,0)
\put(20,20){\color{Turquoise1}\line(1,0){73}}
\multiput(20,20)(.03372164316,.05518087063){1080}{\color{DeepPink1}\line(0,1){.05518087063}}
\put(20,20){\color{DeepPink1}\circle{5.5}}
\put(20,20){\color{DeepPink1}\circle{1.5}}
\put(20,20){\color{Turquoise1}\circle{9}}
\put(56.25,20){\color{Turquoise1}\circle{5.5}}
\put(56.25,20){\color{Turquoise1}\circle{1.5}}
\put(92.5,20){\color{Turquoise1}\circle{5.5}}
\put(92.5,20){\color{Turquoise1}\circle{1.5}}
\put(56.25,79.75){\color{DeepPink1}\circle{5.5}}
\put(56.25,79.75){\color{DeepPink1}\circle{1.5}}
\put(38.75,51.25){\color{DeepPink1}\circle{5.5}}
\put(38.75,51.25){\color{DeepPink1}\circle{1.5}}
\put(15,5){\makebox(0,0)[cc]{$(1,0,0)$}}
\put(52,5){\makebox(0,0)[cc]{$(0,1,1)$}}
\put(92.5,5){\makebox(0,0)[cc]{$(0,1,-1)$}}
\put(28,52.25){\makebox(0,0)[rc]{$(0,1,0)$}}
\put(42.5,80.25){\makebox(0,0)[rc]{$(0,0,1)$}}
\end{picture}
\\
(a)&&(b)
\end{tabular}
\end{center}
\caption{ \label{2009-QvPRtria} Subdiagrams of Figure~\ref{2009-QvPR} allowing (value definite)
chocolate ball realizations.
}
\end{figure}
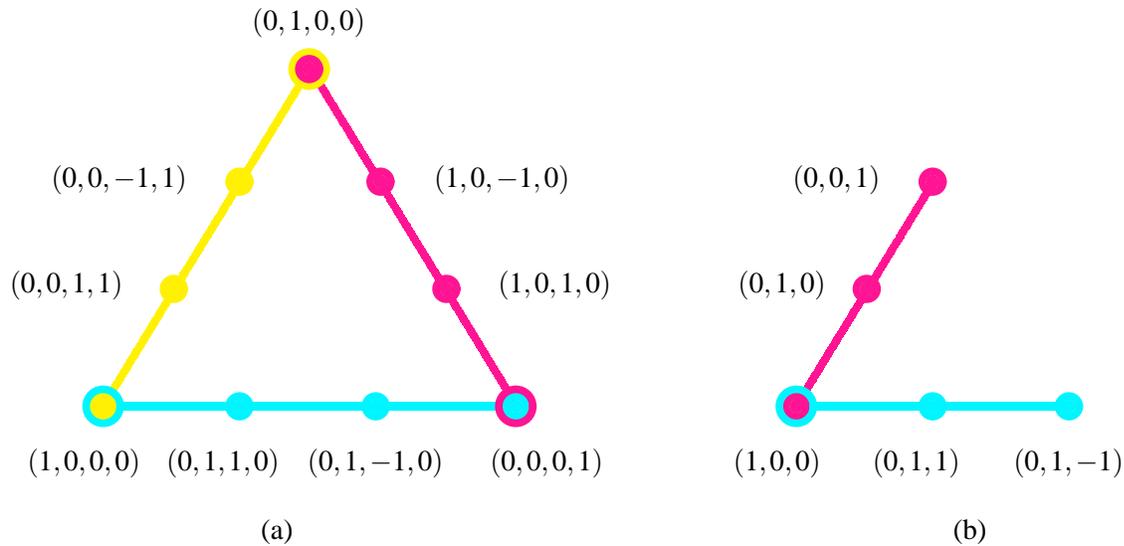

For the sake of a concrete example, we shall consider the tightly interlinked collection of observables
in four-dimensional Hilbert space
presented by Cabello, Estebaranz and Garc{\'i}a-Alcaine~\cite{cabello-96,cabello:210401},
which is depicted in Figure~\ref{2009-QvPR}.
Instead of two measurement bases of two-dimensional Hilbert space used in the BB84 protocol,
nine such bases of four-dimensional Hilbert space, corresponding to the nine smooth (unbroken) orthogonal curves
in Fig.~\ref{2009-QvPR} are used.
In what follows, it is assumed  that any kind of random decision has been prepared
according to the protocol for generating
random sequences sketched above.

\begin{enumerate}
\item
In the first step, ``Alice'' randomly picks an arbitrary basis from the nine available ones,
and sends a random state to ``Bob.''
\item
In the second step, Bob independently from Alice, picks another basis at random,
and measures the particle received from Alice.
\item
In the third step, Alice and Bob compare their bases over a public channel,
and keep only those events which were recorded either in a common basis,
or in an observable interlinking two different bases.
\item
Both then exchange some of the remaining matching outcomes  over a public channel to assure that nobody
has attended their quantum channel.
\item
Bob and Alice encode the four outcomes by four or less different symbols.
As a result, Bob and Alice share a common random key certified by quantum value indefiniteness.
\end{enumerate}

The advantage of this protocol resides in the fact that is does not allow its realization by
any partition of a set, or any kind of colored chocolate balls.
Because if it did, any such coloring could be used to generate ``classical'' two-valued states,
which in turn may be used towards a classical re-interpretation of the quantum observables; an option
ruled out by the Kochen-Specker theorem.

Readers  not  totally convinced at this point might, for the sake of demonstration,
consider a generalized urn model with nine colors, associated with the nine bases
in Figure~\ref{2009-QvPR}.
Suppose further that there is a uniform set of symbols, say $\{0,1,2,3\}$
for all four colors.
If all varieties (permutations) contribute, the number of different types of balls
should be $4^9$. Note, however, that every interlinked color must have {\em identical}
(or at least  unique ``partner'')
symbols in the interlinking colors; a condition which cannot be satisfied
``globally'' for all the interlinks in Figure~\ref{2009-QvPR}.

A simplified version of the protocol, which is based on a subdiagram
of Figure~\ref{2009-QvPR}, contains only three contexts,
which are closely interlinked.
The structure of observables is depicted in Figure~\ref{2009-QvPRtria}(a).
The vectors represent observables in four-dimensional Hilbert space
in their usual interpretation as projectors generating the one-dimensional subspaces spanned by them.
In addition to this quantum mechanical
representation, and in contrast to the Kochen-Specker configuration in Figure~\ref{2009-QvPR},
this global collection of observables
still allows for value definiteness, as there are ``enough'' two valued states
permitting the formation of a partition logic and thus a chocolate ball realization; e.g.,
$$
\begin{array}{c}
\{
\{
\{1,2
\},
\{ 3,4,5,6,7
\},
\{ 8,9,10,11,12
\},
\{13,14
\}
\}, \\
\{
\{1,4,5,9,10
\},
\{ 2,6,7,11,12
\},
\{ 3,8
\},
\{ 13,14
\}
\}, \\
\{
\{ 1,2
\},
\{ 3,8
\},
\{ 4,6,9,11,13
\},
\{ 5,7,10,12,14
\}
\}
\}.
\end{array}
$$
The three partitions of the set $\{1,2,\ldots ,14\}$ have been obtained
by indexing the atoms in terms of
all the nonvanishing
two-valued states on them~\cite{svozil-2001-eua,svozil-2008-ql}, as
depicted in Figure~\ref{2009-qcho-f2vs}.
They can be straightforwardly applied for a chocolate
ball configuration with three colors (say green, red and blue)
and four symbols (say 0, 1, 2, and 3).
The 14 ball types corresponding to the 14 different two-valued measures are as follows:
\unitlength 0.7mm \allinethickness{1pt}\begin{picture}(12,12)\put(6,2){\circle*{12}} \put(6,2){\makebox(0,0)[cc]{${\color{DeepPink1} {\bf 0}}{\color{Turquoise1} {\bf 0}}{\color{yellow} {\bf 0}}$}}\end{picture},
\unitlength 0.7mm \allinethickness{1pt}\begin{picture}(12,12)\put(6,2){\circle*{12}} \put(6,2){\makebox(0,0)[cc]{${\color{DeepPink1} {\bf 0}}{\color{Turquoise1} {\bf 1}}{\color{yellow} {\bf 0}}$}}\end{picture},
\unitlength 0.7mm \allinethickness{1pt}\begin{picture}(12,12)\put(6,2){\circle*{12}} \put(6,2){\makebox(0,0)[cc]{${\color{DeepPink1} {\bf 1}}{\color{Turquoise1} {\bf 2}}{\color{yellow} {\bf 1}}$}}\end{picture},
\unitlength 0.7mm \allinethickness{1pt}\begin{picture}(12,12)\put(6,2){\circle*{12}} \put(6,2){\makebox(0,0)[cc]{${\color{DeepPink1} {\bf 1}}{\color{Turquoise1} {\bf 0}}{\color{yellow} {\bf 2}}$}}\end{picture},
\unitlength 0.7mm \allinethickness{1pt}\begin{picture}(12,12)\put(6,2){\circle*{12}} \put(6,2){\makebox(0,0)[cc]{${\color{DeepPink1} {\bf 1}}{\color{Turquoise1} {\bf 0}}{\color{yellow} {\bf 3}}$}}\end{picture},
\unitlength 0.7mm \allinethickness{1pt}\begin{picture}(12,12)\put(6,2){\circle*{12}} \put(6,2){\makebox(0,0)[cc]{${\color{DeepPink1} {\bf 1}}{\color{Turquoise1} {\bf 1}}{\color{yellow} {\bf 2}}$}}\end{picture},
\unitlength 0.7mm \allinethickness{1pt}\begin{picture}(12,12)\put(6,2){\circle*{12}} \put(6,2){\makebox(0,0)[cc]{${\color{DeepPink1} {\bf 1}}{\color{Turquoise1} {\bf 1}}{\color{yellow} {\bf 3}}$}}\end{picture},
\unitlength 0.7mm \allinethickness{1pt}\begin{picture}(12,12)\put(6,2){\circle*{12}} \put(6,2){\makebox(0,0)[cc]{${\color{DeepPink1} {\bf 2}}{\color{Turquoise1} {\bf 2}}{\color{yellow} {\bf 1}}$}}\end{picture},
\unitlength 0.7mm \allinethickness{1pt}\begin{picture}(12,12)\put(6,2){\circle*{12}} \put(6,2){\makebox(0,0)[cc]{${\color{DeepPink1} {\bf 2}}{\color{Turquoise1} {\bf 0}}{\color{yellow} {\bf 2}}$}}\end{picture},
\unitlength 0.7mm \allinethickness{1pt}\begin{picture}(12,12)\put(6,2){\circle*{12}} \put(6,2){\makebox(0,0)[cc]{${\color{DeepPink1} {\bf 2}}{\color{Turquoise1} {\bf 0}}{\color{yellow} {\bf 3}}$}}\end{picture},
\unitlength 0.7mm \allinethickness{1pt}\begin{picture}(12,12)\put(6,2){\circle*{12}} \put(6,2){\makebox(0,0)[cc]{${\color{DeepPink1} {\bf 2}}{\color{Turquoise1} {\bf 1}}{\color{yellow} {\bf 2}}$}}\end{picture},
\unitlength 0.7mm \allinethickness{1pt}\begin{picture}(12,12)\put(6,2){\circle*{12}} \put(6,2){\makebox(0,0)[cc]{${\color{DeepPink1} {\bf 2}}{\color{Turquoise1} {\bf 1}}{\color{yellow} {\bf 3}}$}}\end{picture},
\unitlength 0.7mm \allinethickness{1pt}\begin{picture}(12,12)\put(6,2){\circle*{12}} \put(6,2){\makebox(0,0)[cc]{${\color{DeepPink1} {\bf 3}}{\color{Turquoise1} {\bf 3}}{\color{yellow} {\bf 2}}$}}\end{picture}, and
\unitlength 0.7mm \allinethickness{1pt}\begin{picture}(12,12)\put(6,2){\circle*{12}} \put(6,2){\makebox(0,0)[cc]{${\color{DeepPink1} {\bf 3}}{\color{Turquoise1} {\bf 3}}{\color{yellow} {\bf 3}}$}}\end{picture}.

Figure~\ref{2009-QvPRtria}(b) contains a three-dimensional subconfiguration
with two complementary contexts interlinked in a single observable.
It again has a value definite representation in terms of partitions of a set,
and thus again a chocolate ball realization with three symbols in two colors; e.g.,
\unitlength 0.7mm \allinethickness{1pt}\begin{picture}(8,8) \put(4,2){\circle*{8}} \put(4,2){\makebox(0,0)[cc]{${\color{DeepPink1} {\bf 0}}{\color{Turquoise1} {\bf 0}}$}} \end{picture},
\unitlength 0.7mm \allinethickness{1pt}\begin{picture}(8,8) \put(4,2){\circle*{8}} \put(4,2){\makebox(0,0)[cc]{${\color{DeepPink1} {\bf 1}}{\color{Turquoise1} {\bf 1}}$}} \end{picture},
\unitlength 0.7mm \allinethickness{1pt}\begin{picture}(8,8) \put(4,2){\circle*{8}} \put(4,2){\makebox(0,0)[cc]{${\color{DeepPink1} {\bf 1}}{\color{Turquoise1} {\bf 2}}$}} \end{picture},
\unitlength 0.7mm \allinethickness{1pt}\begin{picture}(8,8) \put(4,2){\circle*{8}} \put(4,2){\makebox(0,0)[cc]{${\color{DeepPink1} {\bf 2}}{\color{Turquoise1} {\bf 1}}$}} \end{picture}, and
\unitlength 0.7mm \allinethickness{1pt}\begin{picture}(8,8) \put(4,2){\circle*{8}} \put(4,2){\makebox(0,0)[cc]{${\color{DeepPink1} {\bf 2}}{\color{Turquoise1} {\bf 2}}$}} \end{picture}.

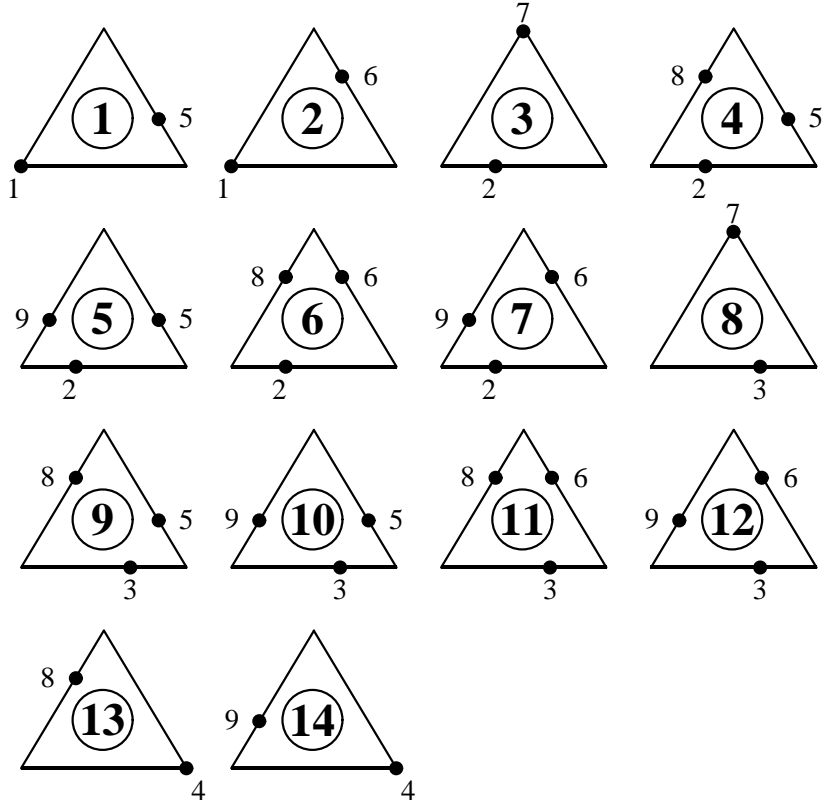
\begin{figure}
\begin{tabular}{cccccccccc}
\unitlength 0.2mm 
\allinethickness{1pt}
\ifx\plotpoint\undefined\newsavebox{\plotpoint}\fi 
\begin{picture}(132.5,122)(0,0)
\put(20,20){\line(1,0){110}}
\put(20,20){\line(3,5){55}}
\put(130,20){\line(-3,5){55}}
\put(20,20){\circle*{8}} \put(15,5){\makebox(0,0)[cc]{$1$}}            
\put(111.25,51.25){\circle*{8}} \put(125,52.25){\makebox(0,0)[lc]{$5$}} 
\put(74,52){\makebox(0,0)[cc]{\Large \bf 1}} \put(74,52){\circle{40}}
\end{picture}
&
\unitlength 0.2mm 
\allinethickness{1pt}
\ifx\plotpoint\undefined\newsavebox{\plotpoint}\fi 
\begin{picture}(132.5,122)(0,0)
\put(20,20){\line(1,0){110}}
\put(20,20){\line(3,5){55}}
\put(130,20){\line(-3,5){55}}
\put(20,20){\circle*{8}} \put(15,5){\makebox(0,0)[cc]{$1$}}            
\put(93.75,79.75){\circle*{8}} \put(108,80.25){\makebox(0,0)[lc]{$6$}}  
\put(74,52){\makebox(0,0)[cc]{\Large \bf 2}} \put(74,52){\circle{40}} \end{picture}
&
\unitlength 0.2mm 
\allinethickness{1pt}
\ifx\plotpoint\undefined\newsavebox{\plotpoint}\fi 
\begin{picture}(132.5,122)(0,0)
\put(20,20){\line(1,0){110}}
\put(20,20){\line(3,5){55}}
\put(130,20){\line(-3,5){55}}
\put(56.25,20){\circle*{8}}\put(52,5){\makebox(0,0)[cc]{$2$}}         
\put(74.75,109.75){\circle*{8}} \put(74.75,122){\makebox(0,0)[cc]{$7$}} 
\put(74,52){\makebox(0,0)[cc]{\Large \bf 3}} \put(74,52){\circle{40}} \end{picture}
&
\unitlength 0.2mm 
\allinethickness{1pt}
\ifx\plotpoint\undefined\newsavebox{\plotpoint}\fi 
\begin{picture}(132.5,122)(0,0)
\put(20,20){\line(1,0){110}}
\put(20,20){\line(3,5){55}}
\put(130,20){\line(-3,5){55}}
\put(56.25,20){\circle*{8}}\put(52,5){\makebox(0,0)[cc]{$2$}}         
\put(111.25,51.25){\circle*{8}} \put(125,52.25){\makebox(0,0)[lc]{$5$}} 
\put(56.25,79.75){\circle*{8}} \put(42.5,80.25){\makebox(0,0)[rc]{$8$}}
\put(74,52){\makebox(0,0)[cc]{\Large \bf 4}} \put(74,52){\circle{40}} \end{picture}
\\
\unitlength 0.2mm 
\allinethickness{1pt}
\ifx\plotpoint\undefined\newsavebox{\plotpoint}\fi 
\begin{picture}(132.5,122)(0,0)
\put(20,20){\line(1,0){110}}
\put(20,20){\line(3,5){55}}
\put(130,20){\line(-3,5){55}}
\put(56.25,20){\circle*{8}}\put(52,5){\makebox(0,0)[cc]{$2$}}         
\put(111.25,51.25){\circle*{8}} \put(125,52.25){\makebox(0,0)[lc]{$5$}} 
\put(38.75,51.25){\circle*{8}} \put(25.5,52.25){\makebox(0,0)[rc]{$9$}}
\put(74,52){\makebox(0,0)[cc]{\Large \bf 5}} \put(74,52){\circle{40}} \end{picture}
&
\unitlength 0.2mm 
\allinethickness{1pt}
\ifx\plotpoint\undefined\newsavebox{\plotpoint}\fi 
\begin{picture}(132.5,122)(0,0)
\put(20,20){\line(1,0){110}}
\put(20,20){\line(3,5){55}}
\put(130,20){\line(-3,5){55}}
\put(56.25,20){\circle*{8}}\put(52,5){\makebox(0,0)[cc]{$2$}}         
\put(93.75,79.75){\circle*{8}} \put(108,80.25){\makebox(0,0)[lc]{$6$}}  
\put(56.25,79.75){\circle*{8}} \put(42.5,80.25){\makebox(0,0)[rc]{$8$}}
\put(74,52){\makebox(0,0)[cc]{\Large \bf 6}} \put(74,52){\circle{40}} \end{picture}
&
\unitlength 0.2mm 
\allinethickness{1pt}
\ifx\plotpoint\undefined\newsavebox{\plotpoint}\fi 
\begin{picture}(132.5,122)(0,0)
\put(20,20){\line(1,0){110}}
\put(20,20){\line(3,5){55}}
\put(130,20){\line(-3,5){55}}
\put(56.25,20){\circle*{8}}\put(52,5){\makebox(0,0)[cc]{$2$}}         
\put(93.75,79.75){\circle*{8}} \put(108,80.25){\makebox(0,0)[lc]{$6$}}  
\put(38.75,51.25){\circle*{8}} \put(25.5,52.25){\makebox(0,0)[rc]{$9$}}
\put(74,52){\makebox(0,0)[cc]{\Large \bf 7}} \put(74,52){\circle{40}} \end{picture}
&
\unitlength 0.2mm 
\allinethickness{1pt}
\ifx\plotpoint\undefined\newsavebox{\plotpoint}\fi 
\begin{picture}(132.5,122)(0,0)
\put(20,20){\line(1,0){110}}
\put(20,20){\line(3,5){55}}
\put(130,20){\line(-3,5){55}}
\put(92.5,20){\circle*{8}}\put(92.5,5){\makebox(0,0)[cc]{$3$}}        
\put(74.75,109.75){\circle*{8}} \put(74.75,122){\makebox(0,0)[cc]{$7$}} 
\put(74,52){\makebox(0,0)[cc]{\Large \bf 8}} \put(74,52){\circle{40}} \end{picture}
\\
\unitlength 0.2mm 
\allinethickness{1pt}
\ifx\plotpoint\undefined\newsavebox{\plotpoint}\fi 
\begin{picture}(132.5,122)(0,0)
\put(20,20){\line(1,0){110}}
\put(20,20){\line(3,5){55}}
\put(130,20){\line(-3,5){55}}
\put(92.5,20){\circle*{8}}\put(92.5,5){\makebox(0,0)[cc]{$3$}}        
\put(111.25,51.25){\circle*{8}} \put(125,52.25){\makebox(0,0)[lc]{$5$}} 
\put(56.25,79.75){\circle*{8}} \put(42.5,80.25){\makebox(0,0)[rc]{$8$}}
\put(74,52){\makebox(0,0)[cc]{\Large \bf 9}} \put(74,52){\circle{40}} \end{picture}
&
\unitlength 0.2mm 
\allinethickness{1pt}
\ifx\plotpoint\undefined\newsavebox{\plotpoint}\fi 
\begin{picture}(132.5,122)(0,0)
\put(20,20){\line(1,0){110}}
\put(20,20){\line(3,5){55}}
\put(130,20){\line(-3,5){55}}
\put(92.5,20){\circle*{8}}\put(92.5,5){\makebox(0,0)[cc]{$3$}}        
\put(111.25,51.25){\circle*{8}} \put(125,52.25){\makebox(0,0)[lc]{$5$}} 
\put(38.75,51.25){\circle*{8}} \put(25.5,52.25){\makebox(0,0)[rc]{$9$}}
\put(74,52){\makebox(0,0)[cc]{\Large \bf 10}} \put(74,52){\circle{40}} \end{picture}
&
\unitlength 0.2mm 
\allinethickness{1pt}
\ifx\plotpoint\undefined\newsavebox{\plotpoint}\fi 
\begin{picture}(132.5,122)(0,0)
\put(20,20){\line(1,0){110}}
\put(20,20){\line(3,5){55}}
\put(130,20){\line(-3,5){55}}
\put(92.5,20){\circle*{8}}\put(92.5,5){\makebox(0,0)[cc]{$3$}}        
\put(93.75,79.75){\circle*{8}} \put(108,80.25){\makebox(0,0)[lc]{$6$}}  
\put(56.25,79.75){\circle*{8}} \put(42.5,80.25){\makebox(0,0)[rc]{$8$}}
\put(74,52){\makebox(0,0)[cc]{\Large \bf 11}} \put(74,52){\circle{40}} \end{picture}
&
\unitlength 0.2mm 
\allinethickness{1pt}
\ifx\plotpoint\undefined\newsavebox{\plotpoint}\fi 
\begin{picture}(132.5,122)(0,0)
\put(20,20){\line(1,0){110}}
\put(20,20){\line(3,5){55}}
\put(130,20){\line(-3,5){55}}
\put(92.5,20){\circle*{8}}\put(92.5,5){\makebox(0,0)[cc]{$3$}}        
\put(93.75,79.75){\circle*{8}} \put(108,80.25){\makebox(0,0)[lc]{$6$}}  
\put(38.75,51.25){\circle*{8}} \put(25.5,52.25){\makebox(0,0)[rc]{$9$}}
\put(74,52){\makebox(0,0)[cc]{\Large \bf 12}} \put(74,52){\circle{40}} \end{picture}
\\
\unitlength 0.2mm 
\allinethickness{1pt}
\ifx\plotpoint\undefined\newsavebox{\plotpoint}\fi 
\begin{picture}(132.5,122)(0,0)
\put(20,20){\line(1,0){110}}
\put(20,20){\line(3,5){55}}
\put(130,20){\line(-3,5){55}}
\put(129.75,20){\circle*{8}} \put(138,5){\makebox(0,0)[cc]{$4$}}      
\put(56.25,79.75){\circle*{8}} \put(42.5,80.25){\makebox(0,0)[rc]{$8$}}
\put(74,52){\makebox(0,0)[cc]{\Large \bf 13}} \put(74,52){\circle{40}} \end{picture}
&
\unitlength 0.2mm 
\allinethickness{1pt}
\ifx\plotpoint\undefined\newsavebox{\plotpoint}\fi 
\begin{picture}(132.5,122)(0,0)
\put(20,20){\line(1,0){110}}
\put(20,20){\line(3,5){55}}
\put(130,20){\line(-3,5){55}}
\put(129.75,20){\circle*{8}} \put(138,5){\makebox(0,0)[cc]{$4$}}      
\put(38.75,51.25){\circle*{8}} \put(25.5,52.25){\makebox(0,0)[rc]{$9$}}
\put(74,52){\makebox(0,0)[cc]{\Large \bf 14}} \put(74,52){\circle{40}} \end{picture}
\end{tabular}
\caption{\label{2009-qcho-f2vs}
Two-valued states interpretable as global truth functions of the observables depicted
in Figure~\ref{2009-QvPRtria}(a). Encircled numbers count the states, smaller numbers
label the observables.
}
\end{figure}

\section{Noncommutative cryptography which cannot be realized quantum mechanically}

\begin{figure}
\begin{center}
\begin{tabular}{cccccccccc}
\unitlength .2mm 
\allinethickness{1pt}
\ifx\plotpoint\undefined\newsavebox{\plotpoint}\fi 
\begin{picture}(130,110)(0,0)
\put(20,20){\line(1,0){110}}
\multiput(20,20)(.1198257081,.1960784314){459}{\line(0,1){.1960784314}}
\multiput(75,110)(.1198257081,-.1960784314){459}{\line(0,-1){.1960784314}}
\put(20,20){\circle*{8}}
\put(15,5){\makebox(0,0)[cc]{$1$}}
\put(102,66.25){\circle*{8}}
\put(111,73){\makebox(0,0)[lc]{$4$}}
\put(74,52){\makebox(0,0)[cc]{\Large \bf 1}}
\put(74,52){\circle{40}}
\end{picture}
&
\unitlength .2mm 
\allinethickness{1pt}
\ifx\plotpoint\undefined\newsavebox{\plotpoint}\fi 
\begin{picture}(138.75,110)(0,0)
\put(20,20){\line(1,0){110}}
\multiput(20,20)(.1198257081,.1960784314){459}{\line(0,1){.1960784314}}
\multiput(75,110)(.1198257081,-.1960784314){459}{\line(0,-1){.1960784314}}
\put(130,20){\circle*{8}}
\put(138.75,5){\makebox(0,0)[cc]{$3$}}
\put(48.25,66.25){\circle*{8}}
\put(30,73){\makebox(0,0)[lc]{$6$}}
\put(74,52){\makebox(0,0)[cc]{\Large \bf 2}}
\put(74,52){\circle{40}}
\end{picture}
&
\unitlength .2mm 
\allinethickness{1pt}
\ifx\plotpoint\undefined\newsavebox{\plotpoint}\fi 
\begin{picture}(130,120.25)(0,0)
\put(20,20){\line(1,0){110}}
\multiput(20,20)(.1198257081,.1960784314){459}{\line(0,1){.1960784314}}
\multiput(75,110)(.1198257081,-.1960784314){459}{\line(0,-1){.1960784314}}
\put(74.25,20){\circle*{8}}
\put(74.25,6){\makebox(0,0)[cc]{$2$}}
\put(75.25,109.75){\circle*{8}}
\put(75.75,123.25){\makebox(0,0)[cc]{$5$}}
\put(74,52){\makebox(0,0)[cc]{\Large \bf 3}}
\put(74,52){\circle{40}}
\end{picture}
&
\unitlength .2mm 
\allinethickness{1pt}
\ifx\plotpoint\undefined\newsavebox{\plotpoint}\fi 
\begin{picture}(130,110)(0,0)
\put(20,20){\line(1,0){110}}
\multiput(20,20)(.1198257081,.1960784314){459}{\line(0,1){.1960784314}}
\multiput(75,110)(.1198257081,-.1960784314){459}{\line(0,-1){.1960784314}}
\put(74.25,20){\circle*{8}}
\put(73,6){\makebox(0,0)[cc]{$2$}}
\put(47.75,66){\circle*{8}}
\put(101.25,66){\circle*{8}}
\put(30,73){\makebox(0,0)[lc]{$6$}}
\put(110.25,73.5){\makebox(0,0)[lc]{$4$}}
\put(74,52){\makebox(0,0)[cc]{\Large \bf 4}}
\put(74,52){\circle{40}}
\end{picture}
\end{tabular}
\end{center}
\caption{ \label{2009-qcho-f2vs-2}Two-valued states on triangular propositional structure with three atoms per context or block.
}
\end{figure}
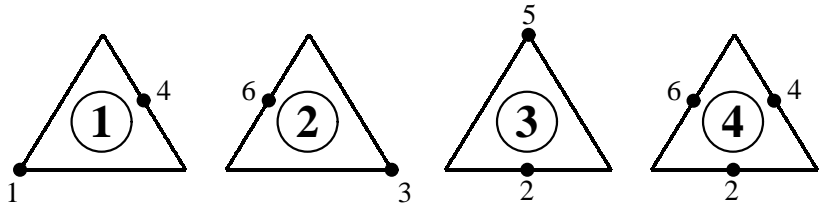

Quantum mechanics does not allow a ``triangular'' structure of observables similar to the one depicted in
Fig.~\ref{2009-QvPRtria} with {\em three} instead of four atoms per block (context),
since no geometric configuration of tripods exist in three-dimensional vector space which would satisfy this scheme.
(For a different propositional structure not satisfiable by quantum mechanics, see Specker's programmatic article~\cite{specker-60} from 1960.)
It contains six atoms $1,\ldots ,6$ in the blocks 1--2--3, 3--4--5, 5--6--1.
In order to obtain a partition logic on which the chocolate ball model can be based, the four two-valued states
are enumerated and depicted in Figure~\ref{2009-qcho-f2vs-2}.

The associated partition logic is given by
$$
\begin{array}{c}
\{
\{
\{1
\},
\{2
\},
\{  3,4
\}
\}, \\
\{
\{1,4
\},
\{ 2
\},
\{ 3
\}
\}, \\
\{
\{ 1
\},
\{ 2,4
\},
\{ 3
\}
\}
\}.
\end{array}
$$
Every one of the three partitions of the set $\{1,\ldots ,4\}$ of ball types labelled by 1 through 4 corresponds to a color; and there are three symbols per colors.
For the first (second/third) partition, the propositions associated with these protocols are:
\begin{itemize}
\item
``when seen through light of the first (second/third) color (e.g., pink/light blue/yellow), symbol ``0'' means ball type number 1 (2/3);''
\item
``when seen through light of the first (second/third) color (e.g., pink/light blue/yellow), symbol ``1'' means ball type number 3 or 4 (1 or 4/2 or 4);''
\item
``when seen through light of the first (second/third) color (e.g., pink/light blue/yellow), symbol ``2'' means ball type number 2 (3/1).''
\end{itemize}
More explicitly, there are four ball types of the form
\unitlength 0.7mm \allinethickness{1pt}\begin{picture}(12,12)\put(6,2){\circle*{12}} \put(6,2){\makebox(0,0)[cc]{${\color{DeepPink1} {\bf 0}}{\color{Turquoise1} {\bf 1}}{\color{yellow} {\bf 2}}$}}\end{picture},
\unitlength 0.7mm \allinethickness{1pt}\begin{picture}(12,12)\put(6,2){\circle*{12}} \put(6,2){\makebox(0,0)[cc]{${\color{DeepPink1} {\bf 2}}{\color{Turquoise1} {\bf 0}}{\color{yellow} {\bf 1}}$}}\end{picture},
\unitlength 0.7mm \allinethickness{1pt}\begin{picture}(12,12)\put(6,2){\circle*{12}} \put(6,2){\makebox(0,0)[cc]{${\color{DeepPink1} {\bf 1}}{\color{Turquoise1} {\bf 2}}{\color{yellow} {\bf 0}}$}}\end{picture}, and
\unitlength 0.7mm \allinethickness{1pt}\begin{picture}(12,12)\put(6,2){\circle*{12}} \put(6,2){\makebox(0,0)[cc]{${\color{DeepPink1} {\bf 1}}{\color{Turquoise1} {\bf 1}}{\color{yellow} {\bf 1}}$}}\end{picture}.
The resulting propositional structure is depicted in Fig.~\ref{2009-QvPRtria-2}.
With respect to realizability,  cryptographic protocols --- such as the one sketched above --- based on this structure are ``stranger than quantum mechanical'' ones.
\begin{figure}
\begin{center}
\unitlength 0.3mm 
\allinethickness{3pt}
\ifx\plotpoint\undefined\newsavebox{\plotpoint}\fi 
\begin{picture}(132.5,122)(0,0)
\put(20,20){\color{Turquoise1}\line(1,0){110}}
\multiput(20,20)(.03372164316,.05518087063){1631}{\color{yellow}\line(0,1){.05518087063}}
\multiput(75,110)(.03372164316,-.05518087063){1631}{\color{DeepPink1}\line(0,-1){.05518087063}}
\put(20,20){\color{yellow}\circle{5.5}}
\put(20,20){\color{yellow}\circle{1.5}}
\put(20,20){\color{Turquoise1}\circle{9}}
\put(74.375,20){\color{Turquoise1}\circle{5.5}}
\put(74.375,20){\color{Turquoise1}\circle{1.5}}
\put(129.75,20){\color{Turquoise1}\circle{5.5}}
\put(129.75,20){\color{Turquoise1}\circle{1.5}}
\put(129.75,20){\color{DeepPink1}\circle{9}}
\put(47.5,65.5){\color{yellow}\circle{5.5}}
\put(47.5,65.5){\color{yellow}\circle{1.5}}
\put(74.75,109.75){\color{yellow}\circle{9}}
\put(74.75,109.75){\color{DeepPink1}\circle{5.5}}
\put(74.75,109.75){\color{DeepPink1}\circle{1.5}}
\put(102.5,65.5){\color{DeepPink1}\circle{5.5}}
\put(102.5,65.5){\color{DeepPink1}\circle{1.5}}
\put(15,5){\makebox(0,0)[cc]{$\{1\}$}}
\put(74.375,5){\makebox(0,0)[cc]{$\{3,4\}$}}
\put(138,5){\makebox(0,0)[cc]{$\{2\}$}}
\put(74.75,122){\makebox(0,0)[cc]{$\{3\}$}}
\put(112,65.8){\makebox(0,0)[lc]{$\{1,4\}$}}
\put(34,65.8){\makebox(0,0)[rc]{$\{2,4\}$}}
\end{picture}
\end{center}
\caption{ \label{2009-QvPRtria-2}Propositional structure allowing (value definite)
chocolate ball realizations with three atoms per context or block which does not allow a quantum analog.
}
\end{figure}
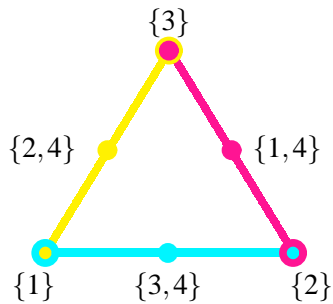

\section{Summary and discussion}

It has been argued that value indefiniteness should be used as a quantum resource
against cryptanalytic attacks, as complementarity may not be a sufficient resource
for the type of ``objective'' security envisaged by quantum cryptography.
A necessary condition for this quantum resource is the presence of at least three
mutually exclusive outcomes.

It may be objected that quantum complementarity suffices
as resource against cryptanalytic attacks,
and thus the original BB84 protocol needs not be amended.
To this criticism I respond with a performance of the original BB84 protocols
with chocolate balls~\cite{svozil-2005-ln1e}; or more formally,
by stating that configurations with just two outcomes leave open the possibility
of a quasi-classical explanation, as they cannot rule out the existence of
sufficiently many two-valued states in order to construct homeomorphisms,
i.e.,  structure-preserving maps between the quantum and classical observables.
Thus, when it comes to fully ``harvesting'' the quantum, it appears prudent to utilize value indefiniteness,
one of its most ``mind-boggling'' features encountered if one assumes the existence of nonoperational yet counterfactual observables.

$\;$\\
{\bf Acknowledgements}
\\
The author gratefully acknowledges discussions with Cristian Calude and Josef Tkadlec, as well as the kind hospitality of the {\it Centre for Discrete Mathematics
and Theoretical Computer Science (CDMTCS)} of the {\it Department of Computer Science at
The University of Auckland.}
This work was also supported by {\it The Department for International Relations}
of the {\em Vienna University of Technology.}
The pink--light blue--yellow coloring scheme is by Renate Bertlmann; communicated to the author by Reinhold Bertlmann.


\end{document}